\documentclass[11pt]{article}
\usepackage{amssymb,amsmath,amsfonts}
\usepackage{graphicx}
\usepackage{graphics}
\usepackage{eepic,epsfig}

\@addtoreset{equation}{section}

\makeatother

\begin{document}

\title{Fermionic condensate in a conical space with\\
a circular boundary and magnetic flux}
\author{S. Bellucci$^{1}$\thanks{%
E-mail: bellucci@lnf.infn.it }, \thinspace\ E. R. Bezerra de Mello$^{2}$%
\thanks{%
E-mail: emello@fisica.ufpb.br}, \thinspace\ A. A. Saharian$^{3}$\thanks{%
E-mail: saharian@ysu.am} \\
\\
\textit{$^1$ INFN, Laboratori Nazionali di Frascati,}\\
\textit{Via Enrico Fermi 40, 00044 Frascati, Italy} \vspace{0.3cm}\\
\textit{$^{2}$Departamento de F\'{\i}sica, Universidade Federal da Para\'{\i}%
ba}\\
\textit{58.059-970, Caixa Postal 5.008, Jo\~{a}o Pessoa, PB, Brazil}\vspace{%
0.3cm}\\
\textit{$^3$Department of Physics, Yerevan State University,}\\
\textit{Alex Manoogian Street, 0025 Yerevan, Armenia}}
\maketitle

\begin{abstract}
The fermionic condensate is investigated in a (2+1)-dimensional conical
spacetime in the presence of a circular boundary and a magnetic flux. It is
assumed that on the boundary the fermionic field obeys the MIT bag boundary
condition. For irregular modes, we consider a special case of boundary
conditions at the cone apex, when the MIT bag boundary condition is imposed
at a finite radius, which is then taken to zero. The fermionic condensate is
a periodic function of the magnetic flux with the period equal to the flux
quantum. For both exterior and interior regions, the fermionic condensate is
decomposed into boundary-free and boundary-induced parts. Two integral
representations are given for the boundary-free part for arbitrary values of
the opening angle of the cone and magnetic flux. At distances from the
boundary larger than the Compton wavelength of the fermion particle, the
condensate decays exponentially with the decay rate depending on the opening
angle of the cone. If the ratio of the magnetic flux to the flux quantum is
not a half-integer number, for a massless field the boundary-free part in
the fermionic condensate vanishes, whereas the boundary-induced part is
negative. For half-integer values of the ratio of the magnetic flux to the
flux quantum, the irregular mode gives non-zero contribution to the
fermionic condensate in the boundary-free conical space.
\end{abstract}

\bigskip


\bigskip

\section{Introduction}

Field theoretical models in 2+1 dimensions exhibit a number of interesting
effects, such as parity violation, flavor symmetry breaking, and
fractionalization of quantum numbers (see Refs. \cite{Dese82}-\cite{Dunn99}%
). An important aspect is the possibility of giving a topological mass to
the gauge bosons without breaking gauge invariance. Field theories in 2+1
dimensions provide simple models in particle physics and related theories
also rise in the long-wavelength description of certain planar condensed
matter systems, including models of high-temperature superconductivity. An
interesting application of Dirac theory in 2+1 dimensions recently appeared
in nanophysics. In a sheet of hexagons from the graphite structure, known as
graphene, the long-wavelength description of the electronic states can be
formulated in terms of the Dirac-like theory of massless spinors in
(2+1)-dimensional spacetime with the Fermi velocity playing the role of
speed of light (for a review see Ref. \cite{Cast09}). One-loop quantum
effects induced by nontrivial topology of graphene made cylindrical and
toroidal nanotubes have been recently considered in Ref. \cite{Bell09}. The
vacuum polarization in graphene with a topological defect is investigated in
Ref. \cite{Site08} within the framework of a long-wavelength continuum model.

The interaction of a magnetic flux tube with a fermionic field gives rise to
a number of interesting phenomena, such as the Aharonov-Bohm effect, parity
anomalies, formation of a condensate and generation of exotic quantum
numbers. For background Minkowski spacetime, the combined effects of the
magnetic flux and boundaries on the vacuum energy have been studied in Refs.
\cite{Lese98,Bene00}. In Ref. \cite{Beze10b} we have investigated the vacuum
expectation value of the fermionic current induced by vortex configuration
of a gauge field in a (2+1)-dimensional conical space with a circular
boundary. On the boundary the fermionic field obeys the MIT bag boundary
condition. Continuing in this line of investigation, in the present paper we
evaluate the fermionic condensate for the same bulk and boundary geometries.
The fermionic condensate is among the most important quantities that
characterize the properties of the quantum vacuum. Although the
corresponding operator is local, due to the global nature of the vacuum,
this quantity carries important information about the global properties of
the background spacetime. The fermionic condensate plays important role in
the models of dynamical breaking of chiral symmetry (see Refs. \cite{Buch89}
for the chiral symmetry breaking in Nambu-Jona-Lasino and Gross-Neveu models
on background of a curved spacetime with non-trivial topology). Note that
the combined effects of the topology and boundaries on the polarization of
the vacuum were studied in Refs. \cite{Brev95}-\cite{Beze08Ferm} for the
cases of scalar, electromagnetic and fermionic fields. In these papers a
cylindrical boundary is considered in the geometry of a cosmic string,
assuming that the boundary is coaxial with the string. The case of a scalar
field was considered in an arbitrary number of spacetime dimensions, whereas
the problems for the electromagnetic and fermionic fields were studied in
four dimensional spacetime. The fermionic condensate in de Sitter spacetime
with toroidally compactified spatial dimensions has been recently
investigated in Refs. \cite{Saha08}.

From the point of view of the physics in the region outside the conical
defect core, the geometry considered in the present paper can be viewed as a
simplified model for the non-trivial core. This model presents a framework
in which the influence of the finite core effects on physical processes in
the vicinity of the conical defect can be investigated. In particular, it
enables us to specify conditions under which the idealized model with the
core of zero thickness can be used. The corresponding results may shed light
upon features of finite core effects in more realistic models, including
those used for defects in crystals and superfluid helium. In addition, the
problem considered here is of interest as an example with combined
topological and boundary-induced quantum effects, in which the vacuum
characteristics can be found in closed analytic form.

The results obtained in the present paper can be applied for the evaluation
of the fermionic condensate in graphitic cones. Graphitic cones are obtained
from the graphene sheet if one or more sectors are excised. The opening
angle of the cone is related to the number of sectors removed, $N_{c}$, by
the formula $2\pi (1-N_{c}/6)$, with $N_{c}=1,2,\ldots ,5$ (for the
electronic properties of graphitic cones see, e.g., \cite{Lamm00} and
references therein). All these angles have been observed in experiments \cite%
{Kris97}. Note that the fermionic condensate in cylindrical and toroidal
carbon nanotubes has been investigated in Ref. \cite{Bell09} within the
framework of the Dirac-like theory for the electronic states in graphene
sheet.

The organization of the paper is as follows. In the next section we evaluate
the fermionic condensate (FC) in a boundary-free conical space with an
infinitesimally thin magnetic flux placed at the apex of the cone. A special
case of boundary conditions at the cone apex is considered, when the MIT bag
boundary condition is imposed at a finite radius, which is then taken to
zero. Two integral representations are provided for the renormalized FC. A
simple expression is found for the special case of the magnetic flux. In
Sect. \ref{sec:FCinside}, we consider the FC in the region inside a circular
boundary with the MIT bag boundary condition. The condensate is decomposed
into boundary-free and boundary-induced parts. Rapidly convergent integral
representation for the latter is obtained. Similar investigation for the
region outside a circular boundary is presented in Sect. \ref{sec:FCoutside}%
. A special case with half-integer values of the ratio of the magnetic flux
to the quantum one is discussed in Sect. \ref{sec:FCspecial}. The main
results are summarized in Sect. \ref{sec:Conc}.

\section{Fermionic condensate in the boundary-free geometry}

\label{sec:BoundFree}

Let us consider a two-component spinor field $\psi $ on background of a $%
(2+1)$-dimensional conical spacetime. The corresponding line element is
given by the expression%
\begin{equation}
ds^{2}=g_{\mu \nu }dx^{\mu }dx^{\nu }=dt^{2}-dr^{2}-r^{2}d\phi ^{2},
\label{ds21}
\end{equation}%
where $r\geqslant 0$, $0\leqslant \phi \leqslant \phi _{0}$, and the points $%
(r,\phi )$ and $(r,\phi +\phi _{0})$ are to be identified. In the discussion
below, in addition to $\phi _{0}$, we use the notation
\begin{equation}
q=2\pi /\phi _{0}.  \label{qu}
\end{equation}%
In the presence of the external electromagnetic field with the vector
potential $A_{\mu }$, the dynamics of the field is governed by the Dirac
equation
\begin{equation}
i\gamma ^{\mu }(\nabla _{\mu }+ieA_{\mu })\psi -m\psi =0\ ,\;\nabla _{\mu
}=\partial _{\mu }+\Gamma _{\mu },  \label{Direq}
\end{equation}%
where $\gamma ^{\mu }=e_{(a)}^{\mu }\gamma ^{(a)}$ are the $2\times 2$ Dirac
matrices in polar coordinates and $e_{(a)}^{\mu }$, $a=0,1,2$, is the basis
tetrad. The operator of the covariant derivative in Eq. (\ref{Direq}) is
defined by the relation
\begin{equation}
\nabla _{\mu }=\partial _{\mu }+\frac{1}{4}\gamma ^{(a)}\gamma
^{(b)}e_{(a)}^{\nu }e_{(b)\nu ;\mu }\ ,  \label{Gammamu}
\end{equation}%
where "$;$" means the standard covariant derivative for vector fields. In $%
(2+1)$-dimensional spacetime there are two inequivalent irreducible
representations of the Clifford algebra. Here we choose the flat space Dirac
matrices in the form $\gamma ^{(0)}=\sigma _{3}$, $\gamma ^{(1)}=i\sigma
_{1} $, $\gamma ^{(2)}=i\sigma _{2}$, with $\sigma _{l}$ being Pauli
matrices. In the second representation the gamma matrices can be taken as $%
\gamma ^{(0)}=-\sigma _{3}$, $\gamma ^{(1)}=-i\sigma _{1}$, $\gamma
^{(2)}=-i\sigma _{2}$. The corresponding results for the second
representation are obtained by changing the sign of the mass, $m\rightarrow
-m$. Note that there is no other $2\times 2$ matrix which anti-commutes with
all $\gamma ^{(a)}$ and, hence, we have no chiral symmetry that would broken
by a mass term in two-dimensional representation.

Our interest in the present paper is the FC, $\langle 0|\bar{\psi}\psi
|0\rangle =\langle \bar{\psi}\psi \rangle $, with $|0\rangle $ being the
vacuum state, in the conical space with a circular boundary. Here and in
what follows $\bar{\psi}=\psi ^{\dagger }\gamma ^{0}$ is the Dirac adjoint
and the dagger denotes Hermitian conjugation. We assume the magnetic field
configuration corresponding to a infinitely thin magnetic flux located at
the apex of the cone. This will be implemented by considering the vector
potential $A_{\mu }=(0,0,A)$ for $r>0$. The quantity $A$ is related to the
magnetic flux $\Phi $ by the formula $A=-\Phi /\phi _{0}$.

First we consider the FC in a boundary-free conical space. It can be
evaluated by using the mode-sum formula
\begin{equation}
\langle \bar{\psi}\psi \rangle =\sum_{\sigma }\bar{\psi}_{\sigma }^{(-)}\psi
_{\sigma }^{(-)},  \label{FCmode}
\end{equation}%
where $\{\psi _{\sigma }^{(+)},\psi _{\sigma }^{(-)}\}$ is a complete set of
positive and negative energy solutions to the Dirac equation specified by
quantum numbers $\sigma $. As it is well known, the theory of von Neumann
deficiency indices leads to a one-parameter family of allowed boundary
conditions in the background of an Aharonov-Bohm gauge field \cite{Sous89}.
Here we consider a special case of boundary conditions at the cone apex,
when the MIT bag boundary condition is imposed at a finite radius, which is
then taken to zero. The FC for other boundary conditions on the cone apex
are evaluated in a way similar to that described below. The contribution of
the regular modes is the same for all boundary conditions and the results
differ by the parts related to the irregular modes.

In the boundary-free conical space the eigenspinors are specified by the set
$\sigma =(\gamma ,j)$ of quantum numbers with $0\leqslant \gamma <\infty $
and $j=\pm 1/2,\pm 3/2,\ldots $. For $j\neq -e\Phi /2\pi $, the
corresponding normalized negative-energy eigenspinors have the form \cite%
{Beze10b}%
\begin{equation}
\psi _{(0)\gamma j}^{(-)}=\left( \gamma \frac{E+m}{2\phi _{0}E}\right)
^{1/2}e^{-iqj\phi +iEt}\left(
\begin{array}{c}
\frac{\gamma \epsilon _{j}e^{-iq\phi /2}}{E+m}J_{\beta _{j}+\epsilon
_{j}}(\gamma r) \\
J_{\beta _{j}}(\gamma r)e^{iq\phi /2}%
\end{array}%
\right) ,  \label{psi0}
\end{equation}%
where $E=\sqrt{\gamma ^{2}+m^{2}}$, $J_{\nu }(x)$ is the Bessel function.
The order of the Bessel function in (\ref{psi0}) is given by the expression%
\begin{equation}
\beta _{j}=q|j+\alpha |-\epsilon _{j}/2,\;q=2\pi /\phi _{0},  \label{jbetj}
\end{equation}%
with%
\begin{equation}
\alpha =eA/q=-e\Phi /2\pi ,  \label{alfatilde}
\end{equation}%
and we have defined%
\begin{equation}
\epsilon _{j}=\left\{
\begin{array}{cc}
1, & \;j>-\alpha \\
-1, & \;j<-\alpha%
\end{array}%
\right. .  \label{epsj}
\end{equation}%
The expression for the positive energy eigenspinor is found from (\ref{psi0}%
) by using the relation $\psi _{\gamma j}^{(+)}=\sigma _{1}\psi _{\gamma
j}^{(-)\ast }$, where the asterisk means complex conjugate. Here we assume
that the parameter $\alpha $ is not a half-integer. The special case of
half-integer $\alpha $ will be considered separately in Sect. \ref%
{sec:FCspecial}.

Substituting the eigenspinors (\ref{psi0}) into the mode-sum (\ref{FCmode}),
for the FC in a boundary-free conical space one finds
\begin{equation}
\langle \bar{\psi}\psi \rangle _{0}=\frac{q}{4\pi }\sum_{j}\int_{0}^{\infty
}d\gamma \frac{\gamma }{E}\left[ (E-m)J_{\beta _{j}+\epsilon
_{j}}^{2}(\gamma r)-(E+m)J_{\beta _{j}}^{2}(\gamma r)\right] ,  \label{FC0}
\end{equation}%
where $\sum_{j}$ means the summation over $j=\pm 1/2,\pm 3/2,\ldots $. Of
course, the expression on the right-hand side of this formula is divergent
and needs to be regularized. We introduce a cutoff function $e^{-s\gamma
^{2}}$ with the cutoff parameter $s>0$. At the end of calculations the limit
$s\rightarrow 0$ is taken. The corresponding regularized expectation value
is presented in the form%
\begin{eqnarray}
&& \langle \bar{\psi}\psi \rangle _{0,\text{reg}} =\frac{q}{4\pi }%
\sum_{j}\int_{0}^{\infty }d\gamma \gamma e^{-s\gamma ^{2}}\left[ J_{\beta
_{j}+\epsilon _{j}}^{2}(\gamma r)-J_{\beta _{j}}^{2}(\gamma r)\right]  \notag
\\
&& \qquad -\frac{qm}{4\pi }\sum_{j}\int_{0}^{\infty }d\gamma \frac{\gamma
e^{-s\gamma ^{2}}}{\sqrt{\gamma ^{2}+m^{2}}}\left[ J_{\beta _{j}+\epsilon
_{j}}^{2}(\gamma r)+J_{\beta _{j}}^{2}(\gamma r)\right] .  \label{FC0reg}
\end{eqnarray}%
The $\gamma $-integral in the first term on the right-hand side is expressed
in terms of the modified Bessel function $I_{\nu }(x)$. In the second term
we use the relation
\begin{equation}
\frac{1}{\sqrt{\gamma ^{2}+m^{2}}}=\frac{2}{\sqrt{\pi }}\int_{0}^{\infty
}dte^{-(\gamma ^{2}+m^{2})t^{2}},  \label{repres}
\end{equation}%
and change the order of integrations. After the evaluation of the $\gamma $%
-integral, the regularized FC is presented in the form:%
\begin{eqnarray}
&& \langle \bar{\psi}\psi \rangle _{0,\text{reg}} =\frac{qe^{-r^{2}/2s}}{%
8\pi s}\sum_{j}[I_{\beta _{j}+\epsilon _{j}}(r^{2}/2s)-I_{\beta
_{j}}(r^{2}/2s)]  \notag \\
&& \qquad -\frac{qme^{m^{2}s}}{2(2\pi )^{3/2}}\sum_{j}\int_{0}^{r^{2}/2s}dx%
\frac{x^{-1/2}e^{-m^{2}r^{2}/2x-x}}{\sqrt{r^{2}-2xs}}[I_{\beta _{j}+\epsilon
_{j}}(x)+I_{\beta _{j}}(x)].  \label{FC0reg2}
\end{eqnarray}%
Before further considering the FC for the general case of the parameters
characterizing the conical structure and the magnetic flux, we study a
special case, which allows us to obtain a simple expression.

\subsection{Special case}

\begin{figure}[tbph]
\begin{center}
\epsfig{figure=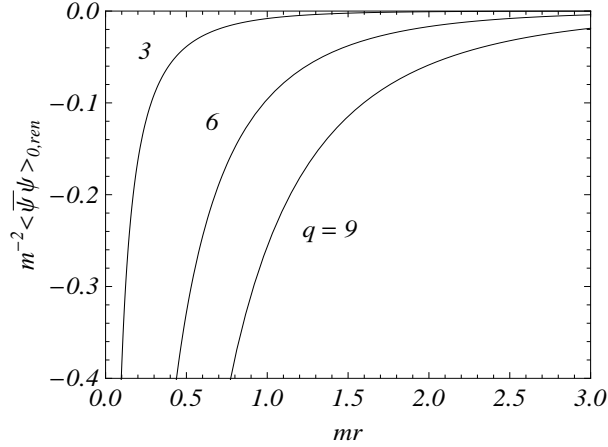,width=8.cm,height=6.cm}
\end{center}
\caption{Fermionic condensate in a boundary-free conical space, as a
function of $mr$ for the special case of integer values of $q$ with the
magnetic flux defined by Eq. (\protect\ref{alphaSpecial}).}
\label{fig1}
\end{figure}

In the special case with $q$ being an integer and
\begin{equation}
\alpha =1/2q-1/2,  \label{alphaSpecial}
\end{equation}%
the orders of the modified Bessel functions in Eq. (\ref{FC0reg2}) become
integer numbers: $\beta _{j}=q|n|$, $j=n+1/2$. The series over $n$ is
summarized explicitly by using the formula \cite{Prud86}%
\begin{equation}
\sideset{}{'}{\sum}_{n=0}^{\infty }I_{qn}(x)=\frac{1}{2q}%
\sum_{k=0}^{q-1}e^{x\cos (2\pi k/q)},  \label{SerSp}
\end{equation}%
where the prime means that the term $n=0$ should be halved. For the
regularized FC we find the expression\footnote{%
Under the condition (\ref{alphaSpecial}), the induced fermionic current in a
higher-dimensional cosmic string spacetime has been analyzed in \cite%
{Beze10c}.}%
\begin{eqnarray}
&& \langle \bar{\psi}\psi \rangle _{0,\text{reg}} =-\frac{1}{4\pi s}%
\sum_{k=1}^{q-1}\sin ^{2}(\pi k/q)e^{-2(r^{2}/2s)\sin ^{2}(\pi k/q)}  \notag
\\
&& \qquad -\frac{me^{m^{2}s}}{(2\pi )^{3/2}}\sum_{k=0}^{q-1}\cos ^{2}(\pi
k/q)\int_{0}^{r^{2}/2s}dx\frac{x^{-1/2}e^{-m^{2}r^{2}/2x}}{\sqrt{r^{2}-2xs}}%
e^{-2x\sin ^{2}(\pi k/q)}.  \label{FC0regSp}
\end{eqnarray}%
The first term on the right-hand side of this formula vanishes in the limit $%
s\rightarrow 0$. In the second term the only divergent contribution in the
limit $s\rightarrow 0$ comes from the $k=0$ term. This term coincides with
the regularized FC in the Minkowski spacetime in the absence of the magnetic
flux. Subtracting this contribution and taking the limit $s\rightarrow 0$,
for the renormalized FC we find%
\begin{equation}
\langle \bar{\psi}\psi \rangle _{0,\text{ren}}=-\frac{m}{4\pi r}%
\sum_{k=1}^{q-1}\frac{\cos ^{2}(\pi k/q)}{\sin (\pi k/q)}e^{-2mr\sin (\pi
k/q)}.  \label{FCren0Sp}
\end{equation}%
Note that the renormalized FC vanishes for a massless field and for a
massive field in a conical space with $q=2$. For other cases the FC is
negative. As expected, it decays exponentially at distances larger that the
Compton wavelength of the fermionic particle. In Fig. \ref{fig1} the FC is
plotted versus $mr$ for different values of $q$. The corresponding values of
the parameter $\alpha $ are found from Eq. (\ref{alphaSpecial}).

\subsection{General case}

For the general case of the parameters $q$ and $\alpha $, as it is seen from
(\ref{FC0reg2}), the regularized FC is expressed in terms of the series%
\begin{equation}
\mathcal{I}(q,\alpha ,z)=\sum_{j}I_{\beta _{j}}(z).  \label{seriesI0}
\end{equation}%
We present the parameter $\alpha $, related to the magnetic flux by Eq. (\ref%
{alfatilde}), in the form%
\begin{equation}
\alpha =\alpha _{0}+n_{0},\;|\alpha _{0}|<1/2,  \label{alf0}
\end{equation}%
with being $n_{0}$ an integer number. Now, Eq. (\ref{seriesI0}) is written as%
\begin{equation}
\mathcal{I}(q,\alpha ,z)=\sum_{n=0}^{\infty }\left[ I_{q(n+\alpha
_{0}+1/2)-1/2}(z)+I_{q(n-\alpha _{0}+1/2)+1/2}(z)\right] ,  \label{seriesI1}
\end{equation}%
which explicitly shows the independence of the series on $n_{0}$. Note that
for the second series appearing in the expression of the FC we have
\begin{equation}
\sum_{j}I_{\beta _{j}+\epsilon _{j}}(z)=\mathcal{I}(q,-\alpha _{0},z).
\label{seriesI2}
\end{equation}%
From these relations we conclude that the FC depends on $\alpha _{0}$ alone
and, hence, it is a periodic function of $\alpha $ with period 1.

In terms of the function (\ref{seriesI0}), the expression (\ref{FC0reg2})
for the regularized FC is written as%
\begin{eqnarray}
&& \langle \bar{\psi}\psi \rangle _{0,\text{reg}} =-\frac{qe^{-r^{2}/2s}}{%
8\pi s}\sum_{\delta =\pm 1}\delta \mathcal{I}(q,\delta \alpha _{0},r^{2}/2s)
\notag \\
&& \qquad -\frac{qme^{m^{2}s}}{2(2\pi )^{3/2}}\int_{0}^{r^{2}/2s}dx\frac{%
x^{-1/2}e^{-m^{2}r^{2}/2x-x}}{\sqrt{r^{2}-2xs}}\sum_{\delta =\pm 1}\mathcal{I%
}(q,\delta \alpha _{0},x).  \label{FC0reg3}
\end{eqnarray}%
For $2p<q<2p+2$ with $p$ being an integer, we use the representation \cite%
{Beze10b}%
\begin{equation}
\mathcal{I}(q,\alpha _{0},z)=\frac{e^{z}}{q}+\mathcal{J}(q,\alpha _{0},z),
\label{CalIJ}
\end{equation}%
with the notation
\begin{eqnarray}
&&\mathcal{J}(q,\alpha _{0},z)=-\frac{1}{\pi }\int_{0}^{\infty }dy\frac{%
e^{-z\cosh y}f(q,\alpha _{0},y)}{\cosh (qy)-\cos (q\pi )}  \notag \\
&&\qquad +\frac{2}{q}\sum_{l=1}^{p}(-1)^{l}\cos [2\pi l(\alpha
_{0}-1/2q)]e^{z\cos (2\pi l/q)}.  \label{seriesI3}
\end{eqnarray}%
The function in the integrand is defined by the expression%
\begin{eqnarray}
f(q,\alpha _{0},y) &=&\cos \left[ q\pi \left( 1/2-\alpha _{0}\right) \right]
\cosh \left[ \left( q\alpha _{0}+q/2-1/2\right) y\right]  \notag \\
&&-\cos \left[ q\pi \left( 1/2+\alpha _{0}\right) \right] \cosh \left[
\left( q\alpha _{0}-q/2-1/2\right) y\right] .  \label{fqualf}
\end{eqnarray}%
In the case $q=2p$, the term
\begin{equation}
-(-1)^{q/2}\frac{e^{-z}}{q}\sin (q\pi \alpha _{0}),  \label{replaced}
\end{equation}%
should be added to the right-hand side of Eq. (\ref{seriesI3}). For $%
1\leqslant q<2$, the last term on the right-hand side of Eq. (\ref{seriesI3}%
) is absent.

In the limit $s\rightarrow 0$, the only divergent contributions to the
functions $e^{-r^{2}/2s}\mathcal{I}(q,\pm \alpha _{0},r^{2}/2s)/s$ come from
the first term in the right-hand side of Eq. (\ref{CalIJ}). The contribution
of this term to the FC does not depend on $\alpha _{0}$ and, consequently,
the divergences are cancelled in the evaluation of the first term in the
right-hand side of (\ref{FC0reg3}). This term vanishes in the limit $%
s\rightarrow 0$ and, hence, it does not contribute to the renormalized FC.
Substituting (\ref{CalIJ}) into the second term in the right-hand side of
Eq. (\ref{FC0reg3}), we see that the only divergent contribution comes from
the term $e^{z}/q$. This contribution does not depend on the opening angle
of the cone and on the magnetic flux. It coincides with the corresponding
quantity in the Minkowski spacetime in the absence of the magnetic flux.
Subtracting the Minkowskian part and taking the limit $s\rightarrow 0$, for
the renormalized FC we find:%
\begin{equation}
\langle \bar{\psi}\psi \rangle _{0,\text{ren}}=-\frac{qm}{2(2\pi )^{3/2}r}%
\int_{0}^{\infty }dx\,x^{-1/2}e^{-m^{2}r^{2}/2x-x}\sum_{\delta =\pm 1}%
\mathcal{J}(q,\delta \alpha _{0},x).  \label{FC0ren1}
\end{equation}%
Note that in the case $q=2p$ the contribution of the additional term (\ref%
{replaced}) to the renormalized FC\ vanishes.

By taking into account Eq. (\ref{seriesI3}), the integration over $x$ in Eq.
(\ref{FC0ren1}) is performed explicitly and one finds the following formula%
\begin{eqnarray}
&&\langle \bar{\psi}\psi \rangle _{0,\text{ren}}=\frac{m}{2\pi r}\Big\{%
-\sum_{l=1}^{p}(-1)^{l}\frac{\cot (\pi l/q)}{e^{2mr\sin (\pi l/q)}}\cos
(2\pi l\alpha _{0})  \notag \\
&&\qquad +\frac{q}{4\pi }\int_{0}^{\infty }dy\frac{e^{-2mr\cosh (y/2)}}{%
\cosh (y/2)}\frac{\sum_{\delta =\pm 1}f(q,\delta \alpha _{0},y)}{\cosh
(qy)-\cos (q\pi )}\Big\},  \label{FC0ren2}
\end{eqnarray}%
where $p$ is an integer defined by $2p\leqslant q<2p+2$. Note that the sum
in the integrand may be written in the form%
\begin{equation}
\sum_{\delta =\pm 1}f(q,\delta \alpha _{0},y)=-2\sinh (y/2)\sum_{\delta =\pm
1}\cos \left[ q\pi \left( 1/2+\delta \alpha _{0}\right) \right] \sinh
[q\left( 1/2-\delta \alpha _{0}\right) y].  \label{fsum}
\end{equation}%
For integer $q$ and for the parameter $\alpha $ given by the special value (%
\ref{alphaSpecial}), from (\ref{FC0ren2}) we obtain the result (\ref%
{FCren0Sp}). At distances larger than the Compton wavelength of the spinor
particle, $mr\gg 1$, the FC is suppressed by the factor $e^{-2mr}$ for $%
1\leqslant q\leqslant 2$ and by the factor $e^{-2mr\sin (\pi /q)}$ for $q>2$%
. In the latter case the main contribution comes from the first term in the
figure braces of the right-hand side in Eq. (\ref{FC0ren2}):%
\begin{equation}
\langle \bar{\psi}\psi \rangle _{0,\text{ren}}\approx \frac{m\cos (2\pi
\alpha _{0})}{2\pi r}\frac{\cot (\pi /q)}{e^{2mr\sin (\pi /q)}},\;mr\gg 1.
\label{FClargem}
\end{equation}

In the special case when the magnetic flux is absent we have $\alpha _{0}=0$
and the general formula (\ref{FC0ren2}) simplifies to%
\begin{eqnarray}
&&\langle \bar{\psi}\psi \rangle _{0,\text{ren}}=-\frac{m}{2\pi r}\Big\{%
\sum_{l=1}^{p}(-1)^{l}\frac{\cot (\pi l/q)}{e^{2mr\sin (\pi l/q)}}  \notag \\
&&\qquad +\frac{2q}{\pi }\cos (q\pi /2)\int_{0}^{\infty }dx\frac{\sinh
\left( qx\right) \tanh (x)e^{-2mr\cosh x}}{\cosh (2qx)-\cos (q\pi )}\Big\}.
\label{FC0renMag0}
\end{eqnarray}%
In this case the FC is only a consequence of the conical structure of the
space. For odd values of the parameter $q$ the second term in the figure
braces vanishes and for the FC we have the simple formula%
\begin{equation}
\langle 0|\bar{\psi}\psi |0\rangle _{0,\text{ren}}=-\frac{m}{2\pi r}%
\sum_{l=1}^{(q-1)/2}(-1)^{l}\frac{\cot (\pi l/q)}{e^{2mr\sin (\pi l/q)}}.
\label{FCoddq}
\end{equation}%
Another limiting case corresponds to the magnetic flux in background of
Minkowski spacetime. In this case, taking $q=1$, from Eq. (\ref{FC0ren2}) we
find%
\begin{equation}
\langle \bar{\psi}\psi \rangle _{0,\text{ren}}=-\frac{m\sin (\pi \alpha _{0})%
}{2\pi ^{2}r}\int_{0}^{\infty }dx\frac{\sinh x}{\cosh ^{2}x}\frac{\sinh
\left( 2\alpha _{0}x\right) }{e^{2mr\cosh x}},  \label{FC0renq1}
\end{equation}%
and the FC is negative for $\alpha _{0}\neq 0$.

\begin{figure}[tbph]
\begin{center}
\begin{tabular}{cc}
\epsfig{figure=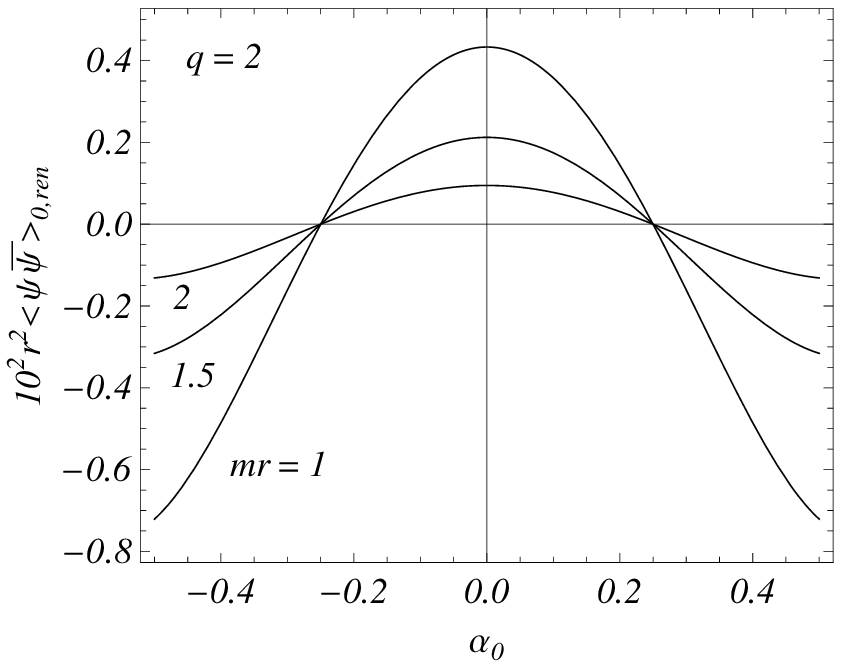,width=7.cm,height=6.cm} & \quad %
\epsfig{figure=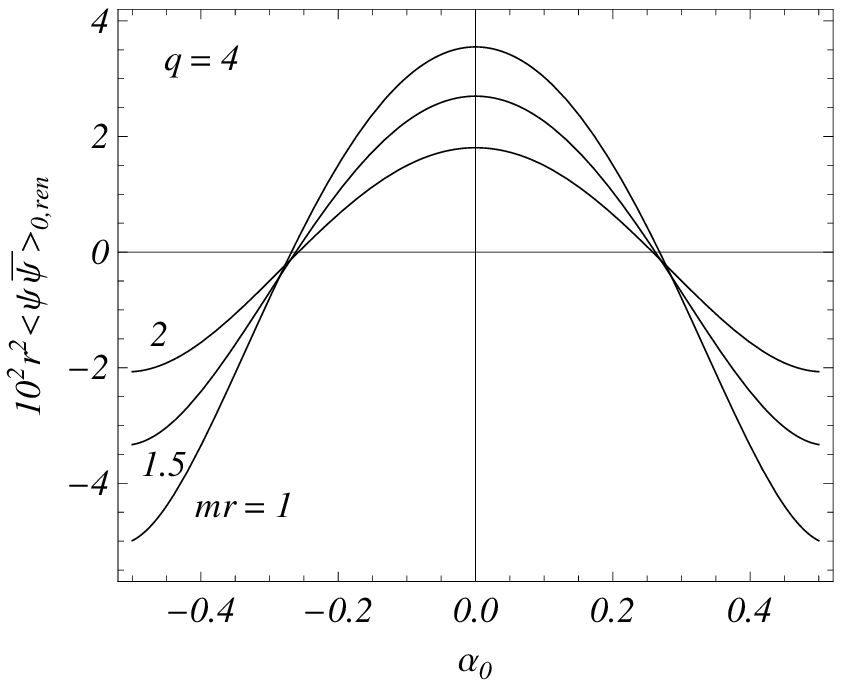,width=7.cm,height=6.cm}%
\end{tabular}%
\end{center}
\caption{The FC as a function of the magnetic flux for a massive fermionic
field in boundary-free conical spaces with $q=2$ (left plot) and $q=4$
(right plot).}
\label{fig2}
\end{figure}

In Fig. \ref{fig2}, the fermionic condensate is plotted as a function of the
magnetic flux for a massive fermionic field in conical spaces with $\phi
_{0}=\pi $ (left plot) and $\phi _{0}=\pi /2$ (right plot). Note that for $%
q=2$ the first term in figure braces of (\ref{FC0ren2}) vanishes and the
second term contains the factor $\cos \left( 2\pi \alpha _{0}\right) $.
Consequently, in this case the FC vanishes at $\alpha _{0}=\pi /4$.

An alternative expression for the FC is obtained by using the formula \cite%
{Beze10b}
\begin{eqnarray}
&&\mathcal{I}(q,\alpha _{0},x)=A(q,\alpha _{0},x)+\frac{2}{q}%
\int_{0}^{\infty }dz\,I_{z}(x)\,  \notag \\
&&\qquad -\frac{4}{\pi q}\int_{0}^{\infty }dz\,{\mathrm{Re}}\left[ \frac{%
\sinh (z\pi )K_{iz}(x)}{e^{2\pi (z+i|q\alpha _{0}-1/2|)/q}+1}\right] ,
\label{Rep2}
\end{eqnarray}%
with $A(q,\alpha _{0},x)=0$ for $|\alpha _{0}-1/2q|\leqslant 1/2$, and
\begin{equation}
A(q,\alpha _{0},x)=\frac{2}{\pi }\sin [\pi (|q\alpha
_{0}-1/2|-q/2)]K_{|q\alpha _{0}-1/2|-q/2}(x),  \label{Aq}
\end{equation}%
for $1/2<|\alpha _{0}-1/2q|<1$. Substituting the representation (\ref{Rep2})
into the expression (\ref{FC0reg3}) for the regularized FC, we see that the
part with the second term on the right-hand side of (\ref{Rep2}) does not
depend on the opening angle of the cone and on the magnetic flux. It is the
same as in the Minkowski bulk in the absence of the magnetic flux and,
hence, it should be subtracted in the renormalization procedure. Subtracting
the part corresponding to $q=1$ and $\alpha _{0}=0$, in the remaining part
the limit $s\rightarrow 0$ can be taken directly. The first term on the
right of Eq. (\ref{FC0reg3}) vanishes in this limit and for the renormalized
FC we find the representation%
\begin{eqnarray}
&&\langle \bar{\psi}\psi \rangle _{0,\text{ren}}=-\frac{2m}{(2\pi )^{5/2}r}%
\int_{0}^{\infty }dx\,x^{-1/2}e^{-m^{2}r^{2}/2x-x}  \notag \\
&&\qquad \times \Big[qB(q(|\alpha _{0}|-1/2)+1/2,x)\,-2\int_{0}^{\infty
}dz\,K_{iz}(x)h(q,\alpha _{0},z)\Big].  \label{FC0ren3}
\end{eqnarray}%
In this formula we have used the notations
\begin{equation}
h(q,\alpha _{0},z)=\sum_{\delta =\pm 1}{\mathrm{Re}}\left[ \frac{\sinh (z\pi
)}{e^{2\pi (z+i|q\delta \alpha _{0}-1/2|)/q}+1}+\frac{\sinh (z\pi )}{e^{2\pi
z}-1}\right] .  \label{hq}
\end{equation}%
and
\begin{equation}
B(y,x)=\left\{
\begin{array}{cc}
0, & y\leqslant 0, \\
\sin (\pi y)K_{y}(x) & y>0.%
\end{array}%
\right.  \label{Byx}
\end{equation}%
The representation (\ref{FC0ren3}) is valid for conical spaces with $q<4$.
For special values $q=2$ and $\alpha _{0}=1/4$, by taking into account that $%
h(2,1/4,z)=0$, we see that the FC defined by (\ref{FC0ren3}) vanishes.

In the case of a magnetic flux in background of the Minkowski spacetime ($%
q=1 $) we find
\begin{eqnarray}
&& \langle \bar{\psi}\psi \rangle _{0,\text{ren}} =-\frac{2m\sin (\pi
|\alpha _{0}|)}{(2\pi )^{5/2}r}\int_{0}^{\infty
}dx\,x^{-1/2}e^{-m^{2}r^{2}/2x-x}  \notag \\
&&\qquad \times \left[ K_{\alpha _{0}}(x)\,-4\sin (\pi |\alpha
_{0}|)\int_{0}^{\infty }dz\,\frac{K_{iz}(x)\cosh (\pi z)}{\cosh (2\pi
z)-\cos (2\pi \alpha _{0})}\right] .  \label{FC0ren3q1}
\end{eqnarray}%
For a conical space in the absence of the magnetic flux the general formula
reduces to%
\begin{eqnarray}
&& \langle \bar{\psi}\psi \rangle _{0,\text{ren}} =\frac{4m}{(2\pi )^{5/2}r}%
\int_{0}^{\infty }dx\,x^{-1/2}e^{-m^{2}r^{2}/2x-x}  \notag \\
&& \qquad \times \int_{0}^{\infty }dz\,K_{iz}(x)\frac{\cosh (\pi z)\cos (\pi
/q)+\cosh [\pi z(2/q-1)]}{\cosh (2\pi z/q)+\cos (\pi /q)}.
\label{FC0ren3alf0}
\end{eqnarray}%
For $q=2$ the integral over $z$ is evaluated explicitly (see, for instance,
\cite{Prud86}) and we get a simple expression $\langle \bar{\psi}\psi
\rangle _{0,\text{ren}}=(m/\pi )^{2}\int_{1}^{\infty }dt\,K_{1}(2mrt)/t$.
Recall that for odd values of $q$ we have the simple formula (\ref{FCoddq}).
For the second representation of the Clifford algebra the renormalized FC in
a boundary-free conical space changes the sign.

We can generalize the results given above for a more general situation where
the spinor field $\psi $ obeys quasiperiodic boundary condition along the
azimuthal direction%
\begin{equation}
\psi (t,r,\phi +\phi _{0})=e^{2\pi i\chi }\psi (t,r,\phi ),  \label{PerBC}
\end{equation}%
with a constant parameter $\chi $, $|\chi |\leqslant 1/2$. With this
condition, the exponential factor in the expression for the eigenspinors (%
\ref{psi0}) has the form $e^{-iq(n+\chi )\phi +iEt}$. The corresponding
expression for the eigenfunctions is obtained from that given above with the
parameter $\alpha $ defined by
\begin{equation}
\alpha =\chi -e\Phi /2\pi .  \label{Replace}
\end{equation}%
The same replacement generalizes the expression of the FC for the case of a
field with periodicity condition (\ref{PerBC}).

In general, the fermionic modes in background of the magnetic vortex are
divided into two classes, regular and irregular (square integrable) ones. In
the problem under consideration, for given $q$ and $\alpha $, the irregular
mode corresponds to the value of $j$ for which $q|j+\alpha |<1/2$. If we
present the parameter $\alpha $ in the form (\ref{alf0}), then the irregular
mode is present if $|\alpha _{0}|>(1-1/q)/2$. This mode corresponds to $%
j=-n_{0}-$sgn$(\alpha _{0})/2$. Note that, in a conical space, under the
condition $|\alpha _{0}|\leqslant (1-1/q)/2$, there are no square integrable
irregular modes. As we have already mentioned, there is a one-parameter
family of allowed boundary conditions for irregular modes. These modes are
parametrized by the angle $\theta $, $0\leqslant \theta <2\pi $ (see Ref.
\cite{Sous89}). For $|\alpha _{0}|<1/2$, the boundary condition, used in
deriving eigenspinors (\ref{psi0}), corresponds to $\theta =3\pi /2$. If $%
\alpha $ is a half-integer, the irregular mode corresponds to $j=-\alpha $
and for the corresponding boundary condition one has $\theta =0$. Note that
in both cases there are no bound states.

\section{Fermionic condensate inside a circular boundary}

\label{sec:FCinside}

In this section we consider the change in the FC induced by a circular
boundary concentric with the apex of the cone. We assume that the field
obeys the MIT bag boundary condition on the circle with radius $a$:
\begin{equation}
\left( 1+in_{\mu }\gamma ^{\mu }\right) \psi \big|_{r=a}=0\ ,  \label{BCMIT}
\end{equation}%
where $n_{\mu }$ is the outward oriented normal (with respect to the region
under consideration) to the boundary. For the interior region $n_{\mu
}=\delta _{\mu }^{1}$. In this region the negative-energy eigenspinors are
given by the expression \cite{Beze10b}
\begin{equation}
\psi _{\gamma j}^{(-)}=\varphi _{0}e^{-iqj\phi +iEt}\left(
\begin{array}{c}
\frac{\epsilon _{j}\gamma e^{-iq\phi /2}}{E+m}J_{\beta _{j}+\epsilon
_{j}}(\gamma r) \\
e^{iq\phi /2}J_{\beta _{j}}(\gamma r)%
\end{array}%
\right) ,  \label{psijInt}
\end{equation}%
with the same notations as in Eq. (\ref{psi0}). From the boundary condition
at $r=a$ we find that the eigenvalues of $\gamma $ are solutions of the
equation%
\begin{equation}
J_{\beta _{j}}(\gamma a)-\frac{\gamma \epsilon _{j}J_{\beta _{j}+\epsilon
_{j}}(\gamma a)}{\sqrt{\gamma ^{2}+m^{2}}+m}=0.  \label{gamVal}
\end{equation}%
For a given $\beta _{j}$, Eq. (\ref{gamVal}) has an infinite number of
solutions which we denote by $\gamma a=\gamma _{\beta _{j},l}$, $%
l=1,2,\ldots $. The normalization coefficient in Eq. (\ref{psijInt}) is
given by the expression%
\begin{equation}
\varphi _{0}^{2}=\frac{yT_{\beta _{j}}(y)}{2\phi _{0}a^{2}}\frac{\mu +\sqrt{%
y^{2}+\mu ^{2}}}{\sqrt{y^{2}+\mu ^{2}}},  \label{phi0T}
\end{equation}%
with the notations $\mu =ma$ and%
\begin{equation}
T_{\beta _{j}}(y)=\frac{y}{J_{\beta _{j}}^{2}(y)}\Big[y^{2}+\left( \mu
-\epsilon _{j}\beta _{j}\right) \left( \mu +\sqrt{y^{2}+\mu ^{2}}\right) -%
\frac{y^{2}}{2\sqrt{y^{2}+\mu ^{2}}}\Big]^{-1}.  \label{Tnu}
\end{equation}

Substituting the eigenspinors (\ref{psijInt}) into the mode-sum formula%
\begin{equation}
\langle \bar{\psi}\psi \rangle =\sum_{j}\sum_{l=1}^{\infty }\,\bar{\psi}%
_{\gamma j}^{(-)}\psi _{\gamma j}^{(-)},  \label{modesumInt}
\end{equation}%
for the FC we find%
\begin{equation}
\langle \bar{\psi}\psi \rangle =\frac{q}{4\pi a^{2}}\sum_{j}\sum_{l=1}^{%
\infty }yT_{\beta _{j}}(y)\Big[\Big(1-\frac{\mu }{\sqrt{y^{2}+\mu ^{2}}}\Big)%
J_{\beta _{j}+\epsilon _{j}}^{2}(yr/a)-\Big(1+\frac{\mu }{\sqrt{y^{2}+\mu
^{2}}}\Big)J_{\beta _{j}}^{2}(yr/a)\Big],  \label{FCInt}
\end{equation}%
with $y=\gamma _{\beta _{j},l}$. Here we assume that a cutoff function is
introduced without explicitly writing it. The specific form of this function
is not important for the discussion below.

For the summation of the series over $l$ in Eq. (\ref{FCInt}) we use the
summation formula (see \cite{Saha04,Saha08Book})%
\begin{eqnarray}
&&\sum_{l=1}^{\infty }f(\gamma _{\beta _{j},l})T_{\beta }(\gamma _{\beta
_{j},l})=\int_{0}^{\infty }dx\,f(x)-\frac{1}{\pi }\int_{0}^{\infty }dx
\notag \\
&&\quad \times \bigg[e^{-\beta _{j}\pi i}f(xe^{\pi i/2})\frac{K_{\beta
_{j}}^{(+)}(x)}{I_{\beta _{j}}^{(+)}(x)}+e^{\beta _{j}\pi i}f(xe^{-\pi i/2})%
\frac{K_{\beta _{j}}^{(+)\ast }(x)}{I_{\beta _{j}}^{(+)\ast }(x)}\bigg],
\label{SumForm}
\end{eqnarray}%
where the asterisk means complex conjugate. In this formula, for a given
function $F(x)$, we use the notation%
\begin{equation}
F^{(+)}(x)=\left\{
\begin{array}{cc}
xF^{\prime }(x)+(\mu +\sqrt{\mu ^{2}-x^{2}}-\epsilon _{j}\beta _{j})F(x), &
x<\mu , \\
xF^{\prime }(x)+\left( \mu +i\sqrt{x^{2}-\mu ^{2}}-\epsilon _{j}\beta
_{j}\right) F(x), & x\geqslant \mu .%
\end{array}%
\right.  \label{F+Int}
\end{equation}%
Note that for $x<\mu $ one has $F^{(+)\ast }(x)=F^{(+)}(x)$. The ratio of
the combinations of the modified Bessel functions in Eq. (\ref{SumForm}) may
be presented in the form%
\begin{equation}
\frac{K_{\beta _{j}}^{(+)}(x)}{I_{\beta _{j}}^{(+)}(x)}=\frac{W_{\beta
_{j},\beta _{j}+\epsilon _{j}}^{(+)}(x)+i\sqrt{1-\mu ^{2}/x^{2}}}{x[I_{\beta
_{j}}^{2}(x)+I_{\beta _{j}+\epsilon _{j}}^{2}(x)]+2\mu I_{\beta
_{j}}(x)I_{\beta _{j}+\epsilon _{j}}(x)},  \label{KIratio}
\end{equation}%
with the notation defined by
\begin{eqnarray}
W_{\beta _{j},\beta _{j}+\epsilon _{j}}^{(\pm )}(x) &=&x\left[ I_{\beta
_{j}}(x)K_{\beta _{j}}(x)-I_{\beta _{j}+\epsilon _{j}}(x)K_{\beta
_{j}+\epsilon _{j}}(x)\right]  \notag \\
&&\pm \mu \left[ I_{\beta _{j}+\epsilon _{j}}(x)K_{\beta _{j}}(x)-I_{\beta
_{j}}(x)K_{\beta _{j}+\epsilon _{j}}(x)\right] .  \label{Wbet}
\end{eqnarray}%
The notation with the lower sign will be used below.

Applying to the series over $l$ in Eq. (\ref{FCInt}) the summation formula
and comparing with Eq. (\ref{FC0}), we see that the term in the FC
corresponding to the first integral in the right-hand side of Eq. (\ref%
{SumForm}) coincides with the condensate in a boundary-free conical space.
As a result, the FC is presented in the decomposed form
\begin{equation}
\langle \bar{\psi}\psi \rangle =\langle \bar{\psi}\psi \rangle _{0,\text{ren}%
}+\langle \bar{\psi}\psi \rangle _{\text{b}},  \label{FCdecomp}
\end{equation}%
where $\langle \bar{\psi}\psi \rangle _{\text{b}}$ is the part induced by
the circular boundary. For the function $f(x)$ corresponding to Eq. (\ref%
{FCInt}), in the second term on the right-hand side of Eq. (\ref{SumForm}),
the part of the integral over the region $(0,\mu )$ vanishes. Consequently,
the boundary-induced contribution for the FC in the region inside the circle
is given by the expression%
\begin{eqnarray}
&&\langle \bar{\psi}\psi \rangle _{\text{b}}=\frac{q}{2\pi ^{2}}%
\sum_{j}\int_{m}^{\infty }dx\,\,x  \notag \\
&&\qquad \times \Big\{m\frac{I_{\beta _{j}}^{2}(xr)-I_{\beta _{j}+\epsilon
_{j}}^{2}(xr)}{\sqrt{x^{2}-m^{2}}}{\mathrm{Re}}[K_{\beta
_{j}}^{(+)}(xa)/I_{\beta _{j}}^{(+)}(xa)]  \notag \\
&&\qquad -[I_{\beta _{j}}^{2}(xr)+I_{\beta _{j}+\epsilon _{j}}^{2}(xr)]{%
\mathrm{Im}}[K_{\beta _{j}}^{(+)}(xa)/I_{\beta _{j}}^{(+)}(xa)]\Big\}.
\label{FCbInt}
\end{eqnarray}%
The real and imaginary parts appearing in this equation are easily obtained
from Eq. (\ref{KIratio}). Note that under the change $\alpha \rightarrow
-\alpha $, $j\rightarrow -j$, we have $\beta _{j}\rightarrow \beta
_{j}+\epsilon _{j}$, $\beta _{j}+\epsilon _{j}\rightarrow \beta _{j}$. From
here it follows that the real/imaginary part in Eq. (\ref{FCbInt}) is an
odd/even function under this change. Now, from Eq. (\ref{FCbInt}) we see
that the boundary-induced part in the FC is an even function of $\alpha $.
For points away from the circular boundary and the cone apex, the
boundary-induced contribution is finite and the renormalization is reduced
to that for the boundary-free geometry. This contribution is a periodic
function of the parameter $\alpha $ with the period equal to 1. So, if we
present this parameter in the form (\ref{alf0}) with $n_{0}$ being an
integer, then the FC depends on $\alpha _{0}$ alone.

In the case of a massless field the expressions for the boundary-induced
part in the FC takes the form%
\begin{equation}
\langle \bar{\psi}\psi \rangle _{\text{b}}=-\frac{q}{2\pi ^{2}a^{2}}%
\sum_{j}\int_{0}^{\infty }dz\,\frac{I_{\beta _{j}}^{2}(zr/a)+I_{\beta
_{j}+\epsilon _{j}}^{2}(zr/a)}{I_{\beta _{j}}^{2}(z)+I_{\beta _{j}+\epsilon
_{j}}^{2}(z)}.  \label{FCbIntm0}
\end{equation}%
As it is seen, this part is always negative. We would like to point out that
the boundary-induced FC does not vanish for a massless filed. The
corresponding boundary-free part vanishes and, hence, for a massless field $%
\langle \bar{\psi}\psi \rangle =\langle \bar{\psi}\psi \rangle _{\text{b}}$.

Various special cases of general formula (\ref{FCbInt}) can be considered.
In the absence of the magnetic flux one has $\alpha =0$ and the
contributions of the negative and positive values of $j$ to the FC coincide.
The corresponding formulas are obtained from (\ref{FCbInt}) and (\ref%
{FCbIntm0}) making the replacements%
\begin{equation}
\sum_{j}\rightarrow 2\sum_{j=1/2,3/2,\ldots },\;\beta _{j}\rightarrow
qj-1/2,\;\beta _{j}+\epsilon _{j}\rightarrow qj+1/2.  \label{ReplaceAlf0}
\end{equation}%
In the case $q=1$, we obtain the FC induced by the magnetic flux
and a circular boundary in the Minkowski spacetime. And finally,
in the simplest case $\alpha =0$ and $q=1$ one has $\langle
\bar{\psi}\psi \rangle _{0,\text{ren}}=0$, and the expression
(\ref{FCbInt}) gives the FC induced by a
circular boundary in the Minkowski bulk:%
\begin{eqnarray}
\langle \bar{\psi}\psi \rangle &=&\frac{1}{\pi ^{2}a^{2}}\sum_{n=0}^{\infty
}\int_{\mu }^{\infty }\frac{dx}{I_{n}^{2}(x)+I_{n+1}^{2}(x)+2\mu
I_{n}(x)I_{n+1}(x)/x}\,  \notag \\
&&\times \Big\{\mu \frac{W_{n,n+1}^{(+)}(x)}{\sqrt{x^{2}-\mu ^{2}}}\left[
I_{n}^{2}(xr/a)-I_{n+1}^{2}(xr/a)\right]  \notag \\
&&-\sqrt{1-\mu ^{2}/x^{2}}\left[ I_{n}^{2}(xr/a)+I_{n+1}^{2}(xr/a)\right] %
\Big\},  \label{FCIntMink}
\end{eqnarray}%
where the function $W_{n,n+1}^{(+)}(x)$ is defined by Eq. (\ref{KIratio}).

Now we turn to the investigation of the FC in asymptotic regions of the
parameters. For large values of the circle radius, we replace the modified
Bessel functions in Eq. (\ref{FCbInt}), with $xa$ in their arguments, by
asymptotic expansions for large values of the argument. In the case of a
massive field the dominant contribution to the integral comes from the
integration range near the lower limit. In the leading order one has%
\begin{equation}
\langle \bar{\psi}\psi \rangle _{\text{b}}\approx \frac{qm^{2}e^{-2ma}}{8%
\sqrt{\pi }(ma)^{3/2}}\sum_{j}\epsilon _{j}\left[ \beta _{j}I_{\beta
_{j}+\epsilon _{j}}^{2}(mr)-(\beta _{j}+\epsilon _{j})I_{\beta _{j}}^{2}(mr)%
\right] ,  \label{FCbIntLargeRad}
\end{equation}%
and for a fixed value of the radial coordinate, the boundary-induced FC is
exponentially small.

For a massless field, assuming $r/a\ll 1$, we expand the modified Bessel
function in the numerator of integrand in Eq. (\ref{FCbIntm0}) in powers of $%
r/a$. The dominant contribution comes from the term $j=1/2$ for $\alpha
_{0}<0$ and from the term $j=-1/2$ for $\alpha _{0}>0$. To the leading order
we find%
\begin{equation}
\langle \bar{\psi}\psi \rangle _{\text{b}}\approx -\frac{q}{2\pi ^{2}a^{2}}%
\frac{(r/2a)^{2q_{\alpha }-1}}{\Gamma ^{2}(q_{\alpha }+1/2)}\int_{0}^{\infty
}dz\,\frac{z^{2q_{\alpha }-1}}{I_{q_{\alpha }+1/2}^{2}(z)+I_{q_{\alpha
}-1/2}^{2}(z)},  \label{FCbIntLargeRadm0}
\end{equation}%
where $q_{\alpha }$ is defined by the relation
\begin{equation}
q_{\alpha }=q(1/2-|\alpha _{0}|).  \label{qalfa}
\end{equation}%
Hence, for a massless field the FC decays as $a^{-(2q_{\alpha }+1)}$.

For points near the apex of the cone, $r\rightarrow 0$, we use the expansion
of the modified Bessel function for small values of the argument. The
leading term in the boundary-induced FC takes the form%
\begin{eqnarray}
&& \langle \bar{\psi}\psi \rangle _{\text{b}} \approx \frac{q}{2\pi ^{2}a^{2}%
}\frac{(r/2a)^{2q_{\alpha }-1}}{\Gamma ^{2}(q_{\alpha }+1/2)}\int_{\mu
}^{\infty }dz\,\frac{z^{2q_{\alpha }}}{\sqrt{z^{2}-\mu ^{2}}}  \notag \\
&& \qquad \times \frac{\mu W_{q_{\alpha }-1/2,q_{\alpha
}+1/2}^{(+)}(z)-(z^{2}-\mu ^{2})/z}{z[I_{q_{\alpha
}-1/2}^{2}(z)+I_{q_{\alpha }+1/2}^{2}(z)]+2\mu I_{q_{\alpha
}-1/2}(z)I_{q_{\alpha }+1/2}(z)}.  \label{j0intApex}
\end{eqnarray}%
Note that for a massless field this expression reduces to Eq. (\ref%
{FCbIntLargeRadm0}). As it is seen, in the limit $r\rightarrow 0$ the
boundary-induced part vanishes when $|\alpha _{0}|<1/2-1/(2q)$ and diverges
for $|\alpha _{0}|>1/2-1/(2q)$. Notice that in the former case the irregular
mode is absent \ and the divergence in the latter case comes from the
irregular mode. For the magnetic vortex in the background Minkowski
spacetime, the boundary-induced contribution diverges as $r^{-2|\alpha
_{0}|} $. In the case $|\alpha _{0}|=1/2-1/(2q)$, corresponding to $%
q_{\alpha }=1/2$, the boundary-induced FC tends to a finite limiting value.

\begin{figure}[tbph]
\begin{center}
\begin{tabular}{cc}
\epsfig{figure=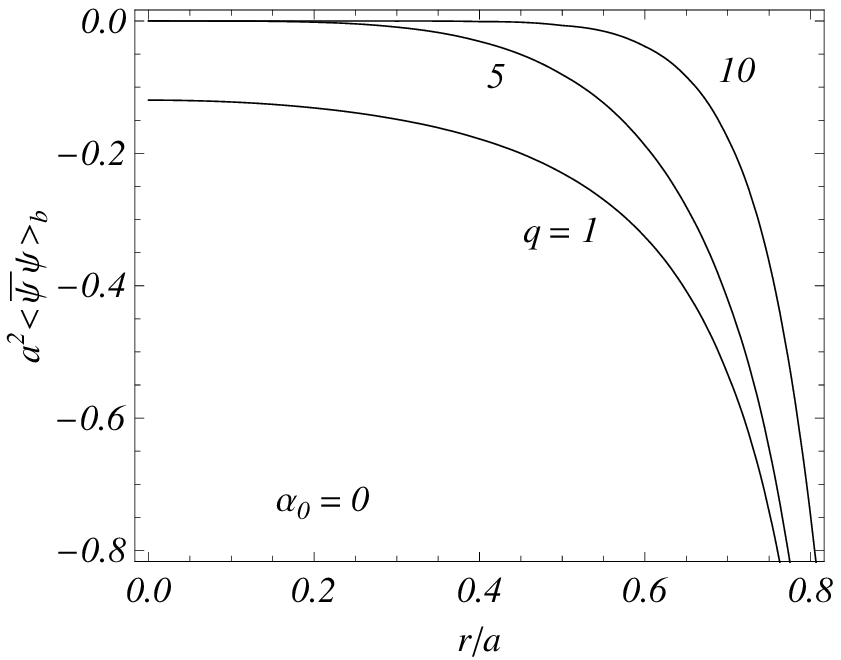,width=7.cm,height=6.cm} & \quad %
\epsfig{figure=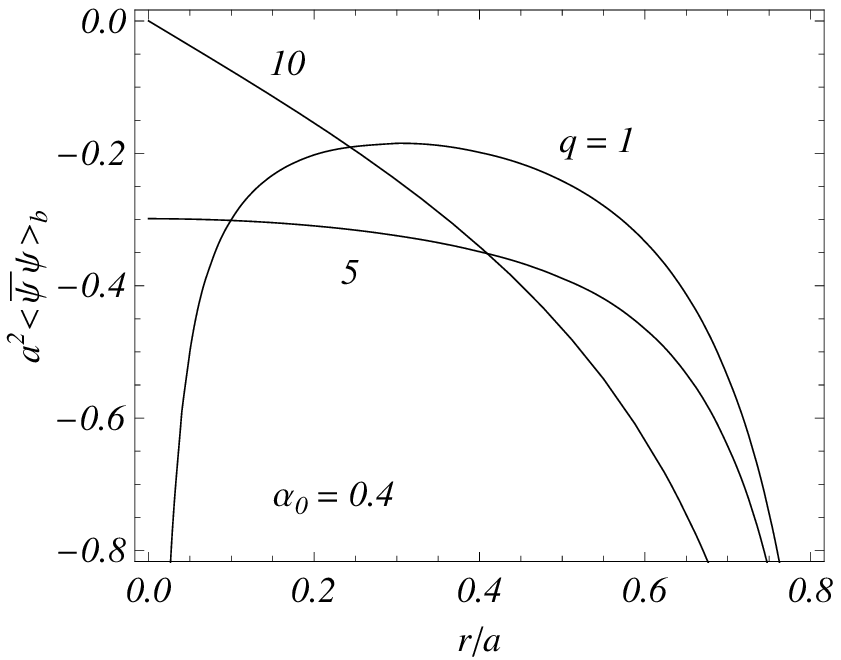,width=7.cm,height=6.cm}%
\end{tabular}%
\end{center}
\caption{The FC inside a circular boundary as a function on the radial
coordinate for a massless fermionic field. }
\label{fig3}
\end{figure}

The boundary-induced part in the FC diverges on the circle. For points near
the circle the main contribution to Eq. (\ref{FCbIntm0}) comes from large
values of $j$. Introducing a new integration variable $y=z/\beta _{j}$, we
use the uniform asymptotic expansion for the modified Bessel function for
large values of the order. To the leading order in the expansion over $%
(1-r/a)$ one finds the behavior%
\begin{equation}
\langle \bar{\psi}\psi \rangle _{\text{b}}\approx -\frac{1}{8\pi (a-r)^{2}}.
\label{FCbIntNear}
\end{equation}%
This leading term does not depend on the opening angle of the cone and on
the magnetic flux. It coincides with the corresponding term for the FC in
the geometry of a circle in (2+1)-dimensional Minkowski spacetime. This
asymptotic behavior is well seen in Fig. \ref{fig3} where the dependence of
the FC on the radial coordinate is presented for a massless fermionic field
for various values of the parameter $q$. The left/right plot corresponds to
the value of the parameter $\alpha _{0}=0$/$\alpha _{0}=0.4$. Note that, in
accordance with the asymptotic analysis given above, for $\alpha _{0}=0.4$
the FC diverges at the cone apex for $q<5$, vanishes for $q>5$ and takes a
finite value for $q=5$. In particular, for $q=10$ one has $\langle \bar{\psi}%
\psi \rangle \propto r$ in the limit $r\rightarrow 0$. These properties are
well seen from the right plot of Fig. \ref{fig3}.

\begin{figure}[tbph]
\begin{center}
\epsfig{figure=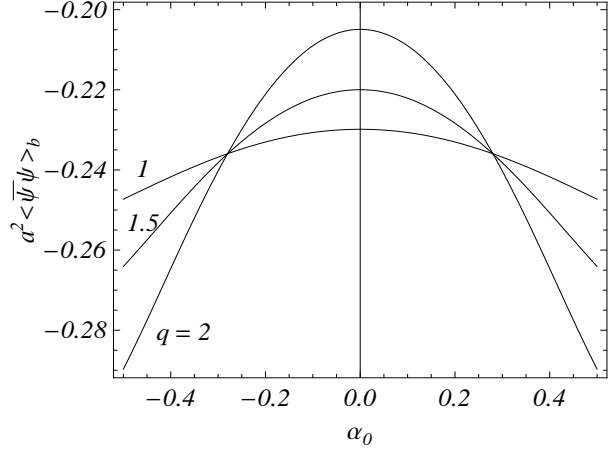,width=8.cm,height=6.cm}
\end{center}
\caption{The FC for a massless field inside a circular boundary as a
function of $\protect\alpha _{0}$.}
\label{fig4}
\end{figure}

In Fig. \ref{fig4}, we present the condensate for a massless fermionic field
inside a circular boundary as a function of the magnetic flux. The graphs
are plotted for $r/a=0.5$ and for several values of the opening angle of the
conical space. Recall that for a massless field the boundary-free part in
the FC vanishes.

\section{Fermionic condensate in the exterior region}

\label{sec:FCoutside}

In the region outside a circular boundary the negative-energy eigenspinors,
obeying the boundary condition (\ref{BCMIT}) with $n_{\mu }=-\delta _{\mu
}^{1}$, have the form \cite{Beze10b}%
\begin{equation}
\psi _{\gamma j}^{(-)}(x)=c_{0}e^{-iqj\phi +iEt}\left(
\begin{array}{c}
\frac{\gamma \epsilon _{j}e^{-iq\phi /2}}{E+m}g_{\beta _{j},\beta
_{j}+\epsilon _{j}}(\gamma a,\gamma r) \\
g_{\beta _{j},\beta _{j}}(\gamma a,\gamma r)e^{iq\phi /2}%
\end{array}%
\right) ,  \label{psiminext}
\end{equation}%
with the function%
\begin{equation}
g_{\nu ,\rho }(x,y)=\bar{Y}_{\nu }^{(-)}(x)J_{\rho }(y)-\bar{J}_{\nu
}^{(-)}(x)Y_{\rho }(y),  \label{gsig}
\end{equation}%
and $Y_{\nu }(x)$\ being the Neumann function. The barred notation in Eq. (%
\ref{gsig}) is defined by the relation%
\begin{equation}
\bar{F}_{\beta _{j}}^{(-)}(z)=-\epsilon _{j}zF_{\beta _{j}+\epsilon
_{j}}(z)-(\sqrt{z^{2}+\mu ^{2}}+\mu )F_{\beta _{j}}(z),  \label{barnot3}
\end{equation}%
with $F=J,Y$ and $\mu =ma$. The normalization coefficient is given by the
expression
\begin{equation}
c_{0}^{2}=\frac{2E\gamma }{\phi _{0}(E+m)}[\bar{J}_{\beta
_{j}}^{(-)2}(\gamma a)+\bar{Y}_{\beta _{j}}^{(-)2}(\gamma a)]^{-1}.
\label{c0}
\end{equation}%
The positive-energy eigenspinors are found with the help of the relation $%
\psi _{\gamma n}^{(+)}=\sigma _{1}\psi _{\gamma n}^{(-)\ast }$. Note that
for the region under consideration the conical singularity is excluded by
the boundary and all modes described by eigenspinors (\ref{psiminext}) are
regular.

Substituting the eigenspinors into the mode-sum formula (\ref{FCmode}), the
FC\ is written in the form%
\begin{equation}
\langle \bar{\psi}\psi \rangle =\frac{q}{4\pi }\sum_{j}\int_{0}^{\infty
}d\gamma \frac{\gamma }{E}\frac{(E-m)g_{\beta _{j},\beta _{j}+\epsilon
_{j}}^{2}(\gamma a,\gamma r)-(E+m)g_{\beta _{j},\beta _{j}}^{2}(\gamma
a,\gamma r)}{\bar{J}_{\beta _{j}}^{(-)2}(\gamma a)+\bar{Y}_{\beta
_{j}}^{(-)2}(\gamma a)}.  \label{FCExt0}
\end{equation}%
As before, we assume the presence of a cutoff function which makes the
expression on the right-hand side of Eq. (\ref{FCExt0}) finite. Similar to
the interior region, the FC\ outside a circular boundary may be written in
the decomposed form (\ref{FCdecomp}).

In order to find an explicit expression for the boundary-induced part, we
note that the boundary-free part is given by Eq. (\ref{FC0}). For the
evaluation of the difference between the total FC and the boundary-free
part, we use the identity%
\begin{equation}
\frac{g_{\beta _{j},\lambda }^{2}(x,y)}{\bar{J}_{\beta _{j}}^{(-)2}(x)+\bar{Y%
}_{\beta _{j}}^{(-)2}(x)}-J_{\lambda }^{2}(y)=-\frac{1}{2}\sum_{l=1,2}\frac{%
\bar{J}_{\beta _{j}}^{(-)}(x)}{\bar{H}_{\beta _{j}}^{(-,l)}(x)}H_{\lambda
}^{(l)2}(y),  \label{ident1}
\end{equation}%
with $\lambda =\beta _{j},\beta _{j}+\epsilon _{j}$, and with $H_{\nu
}^{(l)}(x)$ being the Hankel function. For the boundary-induced part in the
FC we find the expression%
\begin{eqnarray}
&&\langle \bar{\psi}\psi \rangle _{\text{b}}=-\frac{q}{8\pi }%
\sum_{j}\sum_{l=1,2}\int_{0}^{\infty }d\gamma \frac{\gamma }{E}\frac{\bar{J}%
_{\beta _{j}}^{(-)}(\gamma a)}{\bar{H}_{\beta _{j}}^{(-,l)}(\gamma a)}
\notag \\
&&\qquad \times \left[ (E-m)H_{\beta _{j}+\epsilon _{j}}^{(l)2}(\gamma
r)-(E+m)H_{\beta _{j}}^{(l)2}(\gamma r)\right] .  \label{FCExt1}
\end{eqnarray}%
In the complex plane $\gamma $, the integrand of the term with $l=1$ ($l=2$)
decays exponentially in the limit ${\mathrm{Im}}(\gamma )\rightarrow \infty $
[${\mathrm{Im}}(\gamma )\rightarrow -\infty $] for $r>a$ . By using these
properties, we rotate the integration contour in the complex plane $\gamma $
by the angle $\pi /2$ for the term with $l=1$ and by the angle $-\pi /2$ for
the term with $l=2$. The integrals over the segments $(0,im)$ and $(0,-im)$
of the imaginary axis cancel each other. Introducing the modified Bessel
functions, the boundary-induced part in the FC is presented in the form%
\begin{eqnarray}
&&\langle \bar{\psi}\psi \rangle _{\text{b}}=\frac{q}{2\pi ^{2}}%
\sum_{j}\int_{m}^{\infty }dz\,z  \notag \\
&&\quad \times \Big\{m\frac{K_{\beta _{j}}^{2}(zr)-K_{\beta _{j}+\epsilon
_{j}}^{2}(zr)}{\sqrt{z^{2}-m^{2}}}{\mathrm{Re}}[I_{\beta
_{j}}^{(-)}(za)/K_{\beta _{j}}^{(-)}(za)]  \notag \\
&&\quad -[K_{\beta _{j}}^{2}(zr)+K_{\beta _{j}+\epsilon _{j}}^{2}(zr)]{%
\mathrm{Im}}[I_{\beta _{j}}^{(-)}(za)/K_{\beta _{j}}^{(-)}(za)]\Big\},
\label{FCbExt1}
\end{eqnarray}%
where
\begin{equation}
F^{(-)}(z)=zF^{\prime }(z)-(\mu +i\sqrt{z^{2}-\mu ^{2}}+\epsilon _{j}\beta
_{j})F(z).  \label{F+}
\end{equation}

By using the definition (\ref{F+}), the ratio in the integrand of Eq. (\ref%
{FCbExt1}) can be written in the form%
\begin{equation}
\frac{I_{\beta _{j}}^{(-)}(x)}{K_{\beta _{j}}^{(-)}(x)}=\frac{W_{\beta
_{j},\beta _{j}+\epsilon _{j}}^{(-)}(x)+i\sqrt{1-\mu ^{2}/x^{2}}}{x[K_{\beta
_{j}}^{2}(x)+K_{\beta _{j}+\epsilon _{j}}^{2}(x)]+2\mu K_{\beta
_{j}}(x)K_{\beta _{j}+\epsilon _{j}}(x)},  \label{IKratio}
\end{equation}%
with the notation $W_{\beta _{j},\beta _{j}+\epsilon _{j}}^{(-)}(x)$ defined
by Eq. (\ref{Wbet}). Now the real and imaginary parts appearing in Eq. (\ref%
{FCbExt1}) are easily obtained from Eq. (\ref{IKratio}). By taking into
account that under the change $\alpha \rightarrow -\alpha $, $j\rightarrow
-j $, one has $\beta _{j}\rightarrow \beta _{j}+\epsilon _{j}$, $\beta
_{j}+\epsilon _{j}\rightarrow \beta _{j}$, we conclude that the
real/imaginary part in Eq. (\ref{IKratio}) is an odd/even function under
this change. Now, from Eq. (\ref{FCbExt1}) it follows that the
boundary-induced part in the FC is an even function of $\alpha $. This
function is periodic with the period equal to 1.

For a massless field the expression for the boundary-induced part in the FC
simplifies to%
\begin{equation}
\langle \bar{\psi}\psi \rangle _{\text{b}}=-\frac{q}{2\pi ^{2}a^{2}}%
\sum_{j}\int_{0}^{\infty }dz\,\frac{K_{\beta _{j}}^{2}(zr/a)+K_{\beta
_{j}+\epsilon _{j}}^{2}(zr/a)}{K_{\beta _{j}}^{2}(z)+K_{\beta _{j}+\epsilon
_{j}}^{2}(z)}.  \label{FCbExtm0}
\end{equation}%
As in the case of the interior region, the boundary-induced FC does not
vanish for a massless filed. The corresponding boundary-free part vanishes
and, hence, in this case we have $\langle \bar{\psi}\psi \rangle =\langle
\bar{\psi}\psi \rangle _{\text{b}}$. When the magnetic flux is absent, $%
\alpha =0$, the corresponding expression for the boundary-induced part is
obtained from Eq. (\ref{FCbExt1}) by the replacements (\ref{ReplaceAlf0}).
In particular, for the circle in the Minkowski bulk the formula for the
fermionic condensate is obtained from Eq. (\ref{FCIntMink}) by the
interchange $I\rightleftarrows K$, replacing $W_{n,n+1}^{(+)}(x)\rightarrow
W_{n,n+1}^{(-)}(x)$.

Now let us consider the behavior of the boundary-induced part in the FC in
the asymptotic regions of the parameters. First we consider the limit $%
a\rightarrow 0$, for fixed values of $r$. By taking into account the
asymptotics of the modified Bessel functions for small values of the
arguments, to the leading order we find the expression%
\begin{eqnarray}
\langle \bar{\psi}\psi \rangle _{\text{b}} &\approx &\frac{%
q(a/2r)^{2q_{\alpha }}}{\pi ^{2}r^{2}\Gamma ^{2}(q_{\alpha }+1/2)}%
\int_{mr}^{\infty }dz\,\frac{z^{2q_{\alpha }}}{\sqrt{z^{2}-m^{2}r^{2}}}
\notag \\
&&\times \lbrack \left( 2m^{2}r^{2}-z^{2}\right) K_{q_{\alpha
}-1/2}^{2}(z)-z^{2}K_{q_{\alpha }+1/2}^{2}(z)],  \label{FCbExtSma}
\end{eqnarray}%
with the notation (\ref{qalfa}). For a massless field the integral in (\ref%
{FCbExtSma}) is evaluated in terms of the gamma function and one has%
\begin{equation}
\langle \bar{\psi}\psi \rangle _{\text{b}}\approx -\frac{q\Gamma (q_{\alpha
}+1)\Gamma (2q_{\alpha }+1/2)}{2\pi r^{2}\Gamma ^{3}(q_{\alpha }+1/2)}\left(
\frac{a}{2r}\right) ^{2q_{\alpha }}.  \label{FCbExtSmam0}
\end{equation}%
Hence, in the limit $a\rightarrow 0$ and for fixed values of $r$, the
boundary-induced part in FC vanishes as $a^{2q_{\alpha }}$.

\begin{figure}[tbph]
\begin{center}
\begin{tabular}{cc}
\epsfig{figure=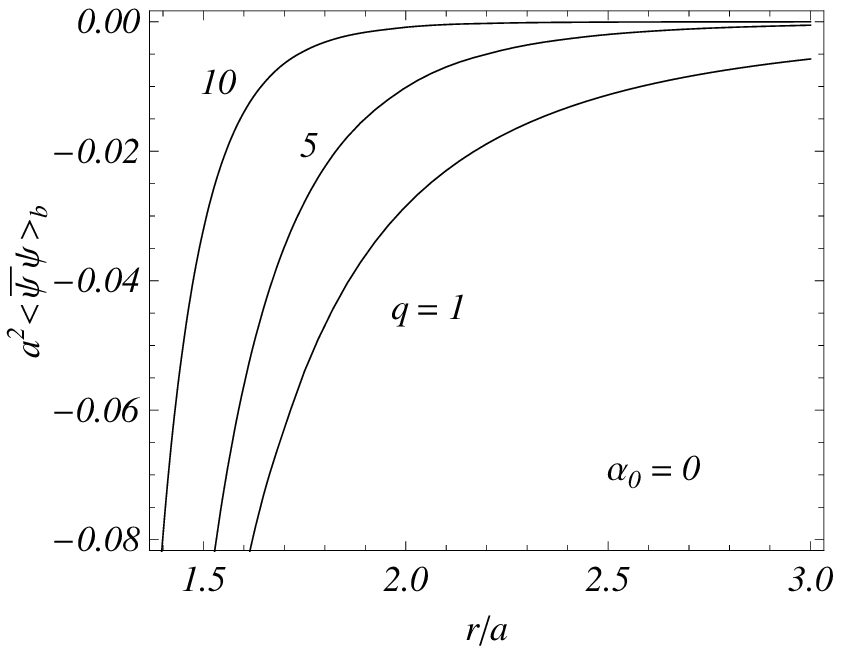,width=7.cm,height=6.cm} & \quad %
\epsfig{figure=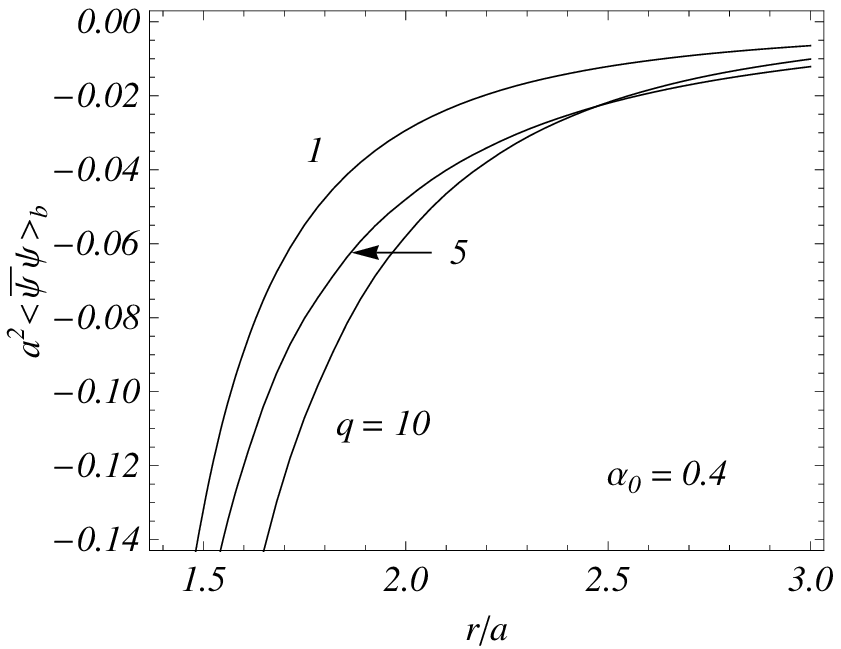,width=7.cm,height=6.cm}%
\end{tabular}%
\end{center}
\caption{The FC outside a circular boundary as a function on the radial
coordinate for a massless fermionic field. }
\label{fig5}
\end{figure}

For a massive field, at large distances from the boundary, under the
condition $mr\gg 1$, the main contribution to the integral in Eq. (\ref%
{FCbExt1}) comes from the region near the lower limit of the integration. In
the leading order we find%
\begin{equation}
\langle \bar{\psi}\psi \rangle _{\text{b}}\approx -\frac{qe^{-2mr}}{4\pi
r^{2}}\sum_{j}{\mathrm{Im}}[I_{\beta _{j}}^{(+)}(ma)/K_{\beta
_{j}}^{(+)}(ma)].  \label{FCExtLarger}
\end{equation}%
and the boundary-induced FC is exponentially suppressed. For a massless
field, the asymptotic at large distances is given by Eq. (\ref{FCbExtSmam0})
and the boundary-induced condensate decays as $r^{-2q_{\alpha }-2}$. For
points near the circle the main contribution to (\ref{FCbExtm0}) comes from
large values of $j$. By using the uniform asymptotic expansion for the
Macdonald function for large values of the order, to the leading order one
finds $\langle \bar{\psi}\psi \rangle _{\text{b}}\approx -[8\pi
(r-a)^{2}]^{-1}$. The leading term in the asymptotic expansion does not
depend on the opening angle of the cone and on the magnetic flux. The
dependence of the FC outside a circular boundary on the radial coordinate is
presented in Fig. \ref{fig5} for a massless field for various values of the
parameter $q$. The left/right plot corresponds to the value of the parameter
$\alpha _{0}=0$/$\alpha _{0}=0.4$.

\begin{figure}[tbph]
\begin{center}
\epsfig{figure=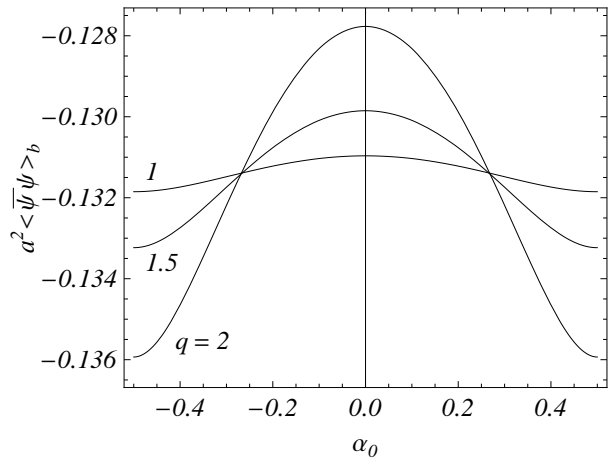,width=8.cm,height=6.cm}
\end{center}
\caption{The FC outside a circular boundary as a function of the magnetic
flux. }
\label{fig6}
\end{figure}

In Fig. \ref{fig6}, the fermionic condensate is plotted for a massless field
outside a circular boundary as a function of the magnetic flux. The graphs
are plotted for $r/a=1.5$ and for several values of the opening angle of the
conical space. For the exterior region there are no irregular modes and the
FC is a continuous function of $\alpha $ at half-integer values. In
particular, its derivative vanishes at these points. Note that this is not
the the case for the interior region.

\section{Half-integer values of the parameter $\protect\alpha $}

\label{sec:FCspecial}

In this section we consider the FC for half-integer values of the parameter $%
\alpha $. In this case for the boundary-free geometry the eigenspinors with $%
j\neq -\alpha $ are still given by Eqs. (\ref{psi0}). For the eigenspinor
corresponding to the special mode with $j=-\alpha $ one has \cite{Beze10b}%
\begin{equation}
\psi _{(0)\gamma ,-\alpha }^{(-)}(x)=\left( \frac{E+m}{\pi \phi _{0}rE}%
\right) ^{1/2}e^{iq\alpha \phi +iEt}\left(
\begin{array}{c}
\frac{\gamma e^{-iq\phi /2}}{E+m}\sin (\gamma r-\gamma _{0}) \\
e^{iq\phi /2}\cos (\gamma r-\gamma _{0})%
\end{array}%
\right) ,  \label{psibetSp}
\end{equation}%
where $\gamma _{0}=\arccos [\sqrt{(E-m)/2E}]$. As we have noted, for
half-integer values of $\alpha $ the mode with $j=-\alpha $ corresponds to
the irregular mode. The contribution of the modes with $j\neq -\alpha $ to
the FC is the same as before. Special consideration is needed for the mode
with $j=-\alpha $ only. For the contribution of this mode to the FC one has%
\begin{eqnarray}
\langle \bar{\psi}\psi \rangle _{0,j=-\alpha } &=&\int_{0}^{\infty }d\gamma
\,\bar{\psi}_{(0)\gamma ,-\alpha }^{(-)}\psi _{(0)\gamma ,-\alpha }^{(-)}
\notag \\
&=&-\frac{q}{2\pi ^{2}r}\int_{0}^{\infty }d\gamma \frac{m+\gamma \sin
(2\gamma r)-m\cos (2\gamma r)}{\sqrt{\gamma ^{2}+m^{2}}}.  \label{FC0HI}
\end{eqnarray}%
The part with the last term in the numerator is finite, whereas the part
with the first two terms is divergent. As before, in order to deal with this
divergence we introduce the cutoff function $e^{-s\gamma ^{2}}$. The
integral in the right-hand side of Eq. (\ref{FC0HI}) is expressed in terms
of the Macdonald function.

For half-integer values of $\alpha $, it can be easily seen that for the
series in the contribution of the modes with $j\neq -\alpha $ one has
\begin{equation}
\sum_{j\neq -\alpha }I_{\beta _{j}}(x)=\sum_{j\neq -\alpha }I_{\beta
_{j}+\epsilon _{j}}(x)=\sum_{n=1}^{\infty }\left[ I_{qn-1/2}(x)+I_{qn+1/2}(x)%
\right] .  \label{SumIbetHI}
\end{equation}%
Summing the contributions from the mode with $j=-\alpha $ and from the modes
$j\neq -\alpha $, for the regularized FC we find the expression%
\begin{eqnarray}
\langle \bar{\psi}\psi \rangle _{0,\text{reg}} &=&-\frac{qme^{m^{2}s}}{(2\pi
)^{3/2}}\sum_{n=1}^{\infty }\int_{0}^{r^{2}/2s}dx\frac{%
x^{-1/2}e^{-m^{2}r^{2}/2x-x}}{\sqrt{r^{2}-2xs}}\left[
I_{qn-1/2}(x)+I_{qn+1/2}(x)\right]  \notag \\
&&-\frac{qm}{4\pi ^{2}r}%
[e^{m^{2}s/2}K_{0}(m^{2}s/2)+2K_{1}(2mr)-2K_{0}(2mr)].  \label{FCregHI}
\end{eqnarray}%
After the summation over $n$ by using the formula given in Sect. \ref%
{sec:BoundFree}, we find the following representation%
\begin{eqnarray}
\langle \bar{\psi}\psi \rangle _{0,\text{reg}} &=&-\frac{m}{2\pi }\left\{
\frac{e^{m^{2}s}}{\sqrt{2\pi }}\int_{0}^{r^{2}/2s}dx\frac{%
x^{-1/2}e^{-m^{2}r^{2}/2x}}{\sqrt{r^{2}-2xs}}+\frac{1}{r}\sum_{l=1}^{p}\frac{%
\cot (\pi l/q)}{e^{2mr\sin (\pi l/q)}}\right.  \notag \\
&&\left. +\frac{q}{2\pi r}\int_{0}^{\infty }dy\frac{\sinh (y/2)\sinh (qy)}{%
\cosh (qy)-\cos (q\pi )}\frac{e^{-2mr\cosh (y/2)}}{\cosh (y/2)}\right\}
\notag \\
&&-\frac{qm}{2\pi ^{2}r}\left[ K_{1}(2mr)-K_{0}(2mr)\right] +o(s),
\label{FCregHI2}
\end{eqnarray}%
where $2p\leqslant q<2p+2$. The first term in the figure braces of this
expression corresponds to the contribution coming from the Minkowski
spacetime part. It is subtracted in the renormalization procedure and for
the renormalized FC in a boundary-free conical space one finds%
\begin{eqnarray}
&&\langle \bar{\psi}\psi \rangle _{0,\text{ren}}=-\frac{qm}{4\pi ^{2}r}%
\int_{0}^{\infty }dy\frac{\sinh (y/2)\sinh (qy)}{\cosh (qy)-\cos (q\pi )}%
\frac{e^{-2mr\cosh (y/2)}}{\cosh (y/2)}  \notag \\
&&\qquad -\frac{m}{2\pi r}\sum_{l=1}^{p}\frac{\cos (\pi l/q)}{\sin (\pi l/q)}%
e^{-2mr\sin (\pi l/q)}-\frac{qm}{2\pi ^{2}r}\left[ K_{1}(2mr)-K_{0}(2mr)%
\right] .  \label{FCrenHI}
\end{eqnarray}%
As before, the FC is a periodic function of $\alpha $ with the period 1.
Note that, in the case under consideration the renormalized FC in a
boundary-free conical space does not vanish for a massless field:%
\begin{equation}
\langle \bar{\psi}\psi \rangle _{0,\text{ren}}=-\frac{q}{4\pi ^{2}r^{2}}%
,\;m=0.  \label{FCrenm0}
\end{equation}%
This corresponds to the contribution of the irregular mode.

Now we consider the region inside a circle with radius $a$. The contribution
of the modes with $j\neq -\alpha $ is given by Eq. (\ref{FCbInt}) where now
the summation goes over $j\neq -\alpha $. For the evaluation of the
contribution coming from the mode with $j=-\alpha $, we note that the
negative-energy eigenspinor for this mode has the form \cite{Beze10b}%
\begin{equation}
\psi _{\gamma ,-\alpha }^{(-)}(x)=\frac{b_{0}}{\sqrt{r}}e^{iq\alpha \phi
+iEt}\left(
\begin{array}{c}
\frac{\gamma e^{-iq\phi /2}}{E+m}\sin (\gamma r-\gamma _{0}) \\
e^{iq\phi /2}\cos (\gamma r-\gamma _{0})%
\end{array}%
\right) ,  \label{psigamSp}
\end{equation}%
where $\gamma _{0}$ is defined after Eq. (\ref{psibetSp}). From boundary
condition (\ref{BCMIT}) it follows that the eigenvalues of $\gamma $ are
solutions of the equation%
\begin{equation}
m\sin (\gamma a)+\gamma \cos (\gamma a)=0.  \label{modeqSp}
\end{equation}%
The positive roots of this equation we denote by $\gamma _{l}=\gamma a$, $%
l=1,2,\ldots $. From the normalization condition, for the coefficient in Eq.
(\ref{psigamSp}) one has%
\begin{equation}
b_{0}^{2}=\frac{E+m}{aE\phi _{0}}\left[ 1-\sin (2\gamma a)/(2\gamma a)\right]
^{-1}.  \label{b02}
\end{equation}

Using Eq. (\ref{psigamSp}), for the contribution of the mode under
consideration to the FC we find:%
\begin{equation}
\langle \bar{\psi}\psi \rangle _{j=-\alpha }=-\frac{1}{a\phi _{0}r}%
\sum_{l=1}^{\infty }\frac{\mu +\gamma _{l}\sin (2\gamma _{l}r/a)-\mu \cos
(2\gamma _{l}r/a)}{\sqrt{\gamma _{l}^{2}+\mu ^{2}}\left[ 1-\sin (2\gamma
_{l})/(2\gamma _{l})\right] },  \label{j02SpAp}
\end{equation}%
where $\mu =ma$ and the presence of a cutoff function is assumed. For the
summation of the series in Eq. (\ref{j02SpAp}), we use the Abel-Plana-type
formula%
\begin{equation}
\sum_{l=1}^{\infty }\frac{\pi f(\gamma _{l})}{1-\sin (2\gamma _{l})/(2\gamma
_{l})}=-\frac{\pi f(0)/2}{1/\mu +1}+\int_{0}^{\infty
}dz\,f(z)-i\int_{0}^{\infty }dz\frac{f(iz)-f(-iz)}{\frac{z+\mu }{z-\mu }%
e^{2z}+1}.  \label{SumFormAp}
\end{equation}%
The latter is obtained from the summation formula given in \cite{Rome02}
(see also \cite{Saha08Book}) taking $b_{1}=0$ and $b_{2}=-1/\mu $. For the
functions $f(z)$ corresponding to Eq. (\ref{j02SpAp}) one has $f(0)=0$. The
second term on the right-hand side of (\ref{SumFormAp}) gives the part
corresponding to the boundary-free geometry. As a result, the FC is
presented in the form
\begin{equation}
\langle \bar{\psi}\psi \rangle _{j=-\alpha }=\langle \bar{\psi}\psi \rangle
_{0,j=-\alpha }+\langle \bar{\psi}\psi \rangle _{\text{b},j=-\alpha },
\label{FCSpIn}
\end{equation}%
where the boundary-induced part is given by the expression%
\begin{equation}
\langle \bar{\psi}\psi \rangle _{\text{b},j=-\alpha }=\frac{q}{\pi ^{2}r}%
\int_{m}^{\infty }dx\frac{m-x\sinh (2xr)-m\cosh (2xr)}{\sqrt{x^{2}-m^{2}}%
\left( \frac{x+m}{x-m}e^{2ax}+1\right) }.  \label{FCbSp}
\end{equation}%
The contribution of the modes $j\neq -\alpha $ remains the same and is
obtained from the corresponding expressions given above for non-half-integer
values of $\alpha $ by the direct substitution $\alpha =1/2$.

Expression (\ref{FCbSp}) for the boundary induced part is simplified for a
massless field%
\begin{equation}
\langle \bar{\psi}\psi \rangle _{\text{b},j=-\alpha }=-\frac{q}{4\pi
^{2}r^{2}}\left[ \frac{\pi r/a}{\sin (\pi r/a)}-1\right] .  \label{FCbSpm0}
\end{equation}%
Note that this part is finite at the circle center. By taking into account
Eq. (\ref{FCrenm0}) and adding the contribution coming from the modes with $%
j\neq -\alpha $, for the total FC one finds%
\begin{equation}
\langle \bar{\psi}\psi \rangle =-\frac{q}{4\pi ar\sin (\pi r/a)}-\frac{q}{%
\pi ^{2}a^{2}}\sum_{n=1}^{\infty }\int_{0}^{\infty }dz\,\frac{%
I_{qn-1/2}^{2}(zr/a)+I_{qn+1/2}^{2}(zr/a)}{%
I_{qn-1/2}^{2}(z)+I_{qn+1/2}^{2}(z)}.  \label{FCtSpm0}
\end{equation}%
The expression on the right-hand side is always negative. The first term
dominates near the cone apex. Near the boundary this term behaves as $%
(1-r/a)^{-1}$, whereas the second term behaves like $(1-r/a)^{-2}$. Hence,
the latter dominates near the circle.

In the region outside a circular boundary there are no irregular modes and
the FC is a continuous function of the parameter $\alpha $ at half-integer
values. The corresponding expression is obtained taking the limit $\alpha
_{0}\rightarrow 1/2$: $\langle \bar{\psi}\psi \rangle =\lim_{\alpha
_{0}\rightarrow 1/2}[\langle \bar{\psi}\psi \rangle _{0,\text{ren}}+\langle
\bar{\psi}\psi \rangle _{\text{b}}]$, where the separate terms are given by
expressions (\ref{FC0ren2}) and (\ref{FCbExt1}). However, note that the
limiting values of the separate terms $\langle \bar{\psi}\psi \rangle _{0,%
\text{ren}}$ and $\langle \bar{\psi}\psi \rangle _{\text{b}}$, defined by
these expressions, do not coincide with the boundary-free and
boundary-induced parts of the FC at half-integer values of $\alpha $.

\section{Conclusion}

\label{sec:Conc}

In this paper we have investigated the FC in a (2+1)-dimensional conical
spacetime with a circular boundary in the presence of a magnetic flux. The
case of massive fermionic field is considered with the MIT bag boundary
condition on the circle. As the first step we have considered a conical
space without boundaries and with a special case of boundary conditions at
the cone apex, when the MIT bag boundary condition is imposed at a finite
radius, which is then taken to zero. For the evaluation of the FC the direct
summation over the modes is used with the spinorial eigenfunctions (\ref%
{psi0}). If the ratio of the magnetic flux to the flux quantum is not a
half-integer number, the regularized FC with the exponential cutoff function
is given by expression (\ref{FC0reg2}). A simple expression for the
renormalized FC, Eq. (\ref{FCren0Sp}), is obtained in the special case when
the parameter $q$ is an integer and is related to the parameter $\alpha $ by
Eq. (\ref{alphaSpecial}). In this special case the renormalized FC vanishes
for a massless field and for a massive field in a conical space with $q=2$
and is negative for other cases.

For the general case of the parameters $q$ and $\alpha $, a convenient
expression for the regularized FC\ is obtained by using the integral
representation (\ref{CalIJ}) for the series involving the modified Bessel
function. This formula allows us to extract explicitly the part in FC
corresponding to the Minkowski spacetime in the absence of the magnetic
flux. Subtracting this part, for the renormalized FC we derived formula (\ref%
{FC0ren2}). At distances larger than the Compton wavelength of the spinor
particle, $mr\gg 1$, the FC is suppressed by the factor $e^{-2mr}$ for $%
1\leqslant q<2$ and by the factor $e^{-2mr\sin (\pi /q)}$ for $q\geqslant 2$%
. In the special case when the magnetic flux is absent the general formula
simplifies to Eq. (\ref{FC0renMag0}). Another limiting case corresponds to
the magnetic flux in background of Minkowski spacetime with the renormalized
FC given by Eq. (\ref{FC0renq1}). An alternative expression for the FC is
obtained by using the integral representation (\ref{Rep2}) for the series
involving the modified Bessel function. This leads to the expression (\ref%
{FC0ren3}) for the renormalized FC. In the special cases of a magnetic flux
in background of the Minkowski spacetime and for a conical space in the
absence of the magnetic flux the general formula reduces to Eqs. (\ref%
{FC0ren3q1}) and (\ref{FC0ren3alf0}), respectively.

In Section \ref{sec:FCinside} we have considered the FC inside a circular
boundary concentric with the apex of the cone. The corresponding
eigenspinors are given by the expression (\ref{psijInt}) and the eigenvalues
of the quantum number $\gamma $ are solutions of Eq. (\ref{gamVal}). The
mode-sum for the FC contains series over these solutions. For the summation
of this series we have used the Abel-Plana-type formula (\ref{SumForm}).
This allows us to decompose the FC into the boundary-free and
boundary-induced parts, Eq. (\ref{FCdecomp}), with the boundary-induced part
given by Eq. (\ref{FCbInt}). The asymptotic near the cone apex is given by
Eq. (\ref{j0intApex}). In this limit the boundary-induced part vanishes when
$|\alpha _{0}|<1/2-1/(2q)$ and diverges for $|\alpha _{0}|>1/2-1/(2q)$. In
the former case the irregular mode is absent \ and the divergence in the
latter case comes from the irregular mode. The boundary-induced FC diverges
on the circle. The leading term in the asymptotic expansion over the
distance from the boundary is given by Eq. (\ref{FCbIntNear}). This term
does not depend on the opening angle of the cone and on the magnetic flux
and coincides with the corresponding term for the FC in the geometry of a
circle in (2+1)-dimensional Minkowski spacetime.

The region outside a circular boundary is considered in Section \ref%
{sec:FCoutside}. The boundary-induced part of the FC\ in this region is
given by Eq. (\ref{FCbExt1}). This expression is obtained from the
corresponding formula for the interior region by the interchange of the
modified Bessel functions $I$ and $K$. For a massless field the general
formula is simplified to Eq. (\ref{FCbExtm0}) and the boundary-induced part
is negative. In the limit when the circle radius tends to zero, $%
a\rightarrow 0$, and for a fixed value of $r$, the boundary-induced part in
FC vanishes as $a^{2q_{\alpha }}$. At large distances from the boundary, for
a massive field, the asymptotic behavior is given by Eq. (\ref{FCExtLarger})
and the boundary-induced FC is exponentially suppressed. For a massless
field, the asymptotic at large distances is given by Eq. (\ref{FCbExtSmam0})
and the boundary-induced condensate decays as $r^{-2q_{\alpha }-2}$.

The special case of the magnetic flux corresponding to half-integer values
of the parameter $\alpha $ is discussed in Section \ref{sec:FCspecial}. For
this case the contribution of the mode with $j=-\alpha $ should be
considered separately. The renormalized FC in the boundary-free geometry is
given by Eq. (\ref{FCrenHI}) and does not vanish in the massless limit. In
the region inside a circular boundary the contribution of the special mode
with $j=-\alpha $ to the FC is given by Eq. (\ref{FCbSp}) and is finite at
the circle center. For a massless fermionic field the total FC inside a
circular boundary is given by Eq. (\ref{FCtSpm0}) and is negative. In the
region outside a circular boundary the FC is a continuous function of the
parameter $\alpha $ at half-integer values and the corresponding expression
is obtained from that in Section \ref{sec:FCoutside} taking the limit $%
\alpha _{0}\rightarrow 1/2$.

\section*{Acknowledgments}

E.R.B.M. thanks Conselho Nacional de Desenvolvimento
Cient\'{\i}fico e Tecnol\'{o}gico (CNPq) for partial financial
support. A.A.S. would like to acknowledge the hospitality of the
INFN Laboratori Nazionali di Frascati, Frascati, Italy.

\end{document}